%Paper: hep-th/9509078
%From: avramidi@math-inf.uni-greifswald.d400.de
%Date: Thu, 14 Sep 1995 11:33:38 +0200

------------------------------ Start of body part 2

%
%
% 11 pages, Plain TeX, 29 KB, no figures
%

\magnification=1200

%\baselineskip=18pt

%
% my conventions
%

\def\h#1{{\cal #1}}

\def\g{\gamma}
\def\d{\delta}
\def\eps{\varepsilon}

\def\m{\mu}
\def\n{\nu}
\def\p{\pi}

\def\s{\sigma}

\def\hb{\hfill\break}

\def\ltextindent#1{\hbox to \hangindent{#1\hss}\ignorespaces}

%%%%%%  expanding to the current date %%%%%%%%%%%%%%%%%%%%%%%%%%%%%%%%%%%%%%
\def\today{\ifcase\month\or
            January\or February\or March\or April\or May\or June\or July\or
            August\or September\or October \or November\or December
            \fi\space\number\day, \number\year
           }

\def\square#1{\mathop{\mkern0.5\thinmuskip
                      \vbox{\hrule
                      \hbox{\vrule
                            \hskip#1
                            \vrule height#1 width 0pt
                            \vrule}
                       \hrule}
                       \mkern0.5\thinmuskip}}
\def\Square{\mathchoice{\square{6pt}}
                       {\square{5pt}}
                       {\square{4pt}}
                       {\square{3pt}}}

% end of definitions --------------------------------

{\nopagenumbers
\null
{\parindent=5cm
{published in Russian in:}

{Yadernaya Fizika, 56 (1993) 245-252}
\bigskip
{translated in English in:}

{Soviet Journal of Nuclear Physics, vol. 56, No 1, (1993) }}

\bigskip
\vskip 6cm
\centerline{\bf A METHOD FOR CALCULATING THE HEAT KERNEL }
\centerline{\bf FOR MANIFOLDS WITH BOUNDARY }
	   \bigskip \bigskip
	  \centerline{I. G. Avramidi}
	   \bigskip
\centerline{Nuclear Physics Department,
Research Institute for Physics, Rostov State University}
	  \centerline{Stachki 194, Rostov-on-Don 344104, Russia}
	   \bigskip
	   \vfill
\centerline{}
\medskip
The covariant technique for calculating the heat kernel asymptotic expansion
for an elliptic differential second order operator  is generalized  to
manifolds with boundary. The first boundary coefficients  of the asymptotic
expansion which are proportional  to $t^{1/2}$ and  $t^{3/2}$
are calculated. Our
results coincide with completely independent results of  previous authors.
	 \vfill
	\eject
}
	\centerline{\bf	 1. Introduction}
	\bigskip
One of the most fruitful approaches in quantum field theory, especially in
gauge theories and quantum gravity, is the approach of the effective action.
It is applicable for topologically nontrivial manifolds (noncomplete
manifolds with boundary etc.) as well.
 The effective action is calculable only
within the limits of some perturbation theory.
In general case of arbitrary background it is impossible to get an exact
answer even in one-loop approximation.
That is why one should use various approximate schemes for the investigation
of various aspects of the effective action. One of the most important
features of such calculational schemes should be manifest covariance of
the calculations at each order.
\par
The general framework of covariant methods for calculating the effective
action is the heat kernel method [1]. Heat kernel plays very important role
in quantum field theory. It has been successfully applied to the analysis of
the structure of ultraviolet divergences and anomalies and renormalization
[2] as well as to calculation of the finite part of the effective action that
is expressible finitely in terms of the coefficients of the heat
kernel asymptotic expansion.
\par
Various methods for calculating heat kernel asymptotic expansion were
developed [3-5]. However, most of them come to nothing more than manifolds
without boundaries and can not be applied immediately to boundary problems.
\par
Recently some papers dealing with boundaries appeared.
These are, mainly, indirect methods  that  use   the factorization
properties of heat kernel and its behaviour under conformal transformations
[6-8] or  an expansion of the boundary in the neighbourhood of the tangent
plane in some point [9,10].
On the other hand the general methods used in mathematical literature [11]
are applicable in any case but the general covariance is lost and the method
becomes not effective at higher orders. Nevertheless, the most complete
summary of formulae concerning boundary contribution in heat kernel
asymptotic expansion is presented in [6].
	\par
In this paper we propose a new algorithm for calculating boundary
contributions in heat kernel asymptotic expansion. We show how the ordinary
technique [1,4] can be applied for manifolds with boundary. A very close
approach was developed in [12].
\bigskip
\bigskip
\bigskip
\centerline{\bf 2. The general framework of calculations }
	\bigskip
	Let $M$ be a $d$-dimensional compact riemannian manifold with smooth
	boundary $\partial M$, and $F$ be elliptic second order differential
	operator on $M$ of the form
$$
F=-\Square + Q         \qquad , \qquad
	\Square =g^{\mu\nu}\nabla_\mu\nabla_\nu
						       \eqno (1)
   $$
  where $\nabla$ is a connection on smooth vector  bundle $V$ over
$M$  and $Q$ is an endomorphism of this bundle.
	\par
The heat kernel is defined by the equation
$$
\left({\partial\over\partial t}+F\right)  U(t\vert x,x^\prime)=0 \  \ ,
                                                                \eqno (2)
$$
with initial condition
$$
U(0\vert x,x^\prime) = g^{-1/2}(x)\delta (x,x^ \prime)
                                                                  \eqno (3)
$$
	\par
In the case of manifolds with boundary one should also impose suitable
boundary conditions:
	$$
	BU(t\vert x,x')\bigg\vert _{x \in \partial M}=0            \eqno (4)
	$$
Dirichlet
	$$
	B=1                                  \eqno (5)
	$$
or Neumann ones
	$$
	B=(n^\m \nabla_\m + S)               \eqno (6)
	$$
where $n^\m (x)$ is the unit inward pointing normal to the
boundary and  $S$ is an endomorphism of the bundle $V$.
	\par
In the case of compact manifolds without boundary  conditions of
periodicity are used instead of the boundary conditions (4). One can show
that for any  $t>0$
the heat kernel is a smooth analytic function on the manifold $M$,
that behaves like a distribution ($\d$-function) near the boundary.
	  \par
In this paper we shall investigate the heat kernel, mainly, when the points
	   $x$  and $x'$ are close together, as it is the limit
 $x \to x' $ which is of interest in quantum field theory
for calculating the effective action and vacuum expectation values
of local observables.
	  \par
Having in mind the calculation of heat kernel asymptotic expansion
at $t\to 0$ let us take advantage of the quasi-classical approximation.
Let us represent the solution of the equation (2) in the form
$$
U(t\vert x,x')=(4\pi t)^{-d/2}\Delta^{1/2}(x,x')
\exp\left(-{\sigma(x,x')\over 2t}\right)\ \Omega (t\vert x,x')
                                                                     \eqno(7)
 $$
where
 $ \sigma (x,x')$
     is a symmetric biscalar satisfying the equation
        $$
        \sigma ={1\over 2}\sigma _\mu\sigma^\mu=
        {1\over 2}\sigma_{\mu^\prime}\sigma^{\mu^\prime}     \qquad ,
                                                            \eqno (8)
        $$
        $$
        \sigma_\mu=\nabla_\mu\sigma\ \ \ ,\ \ \
        \sigma_{\mu'}=\nabla_{\mu'}\sigma
        $$
	and another biscalar $\Delta (x,x')$ is defined by
       $$
       \Delta (x,x')=g^{-1/2}(x)\det
       \left (-\nabla_{\mu'}\nabla_\nu\sigma(x,x')\right)g^{-1/2}(x')
                                                                 \eqno (9)
       $$
	and satisfies, as it is easy to show from (8), the equation
       $$
       \Delta^{-1/2}D\Delta^{1/2}={1\over 2}(d-\Square \sigma)
                                                                 \eqno (10)
       $$
	\par
Making use of (2) and  (7)-(10) we obtain a transfer equation
for the function  $\Omega(t\vert x,x')$
$$
\left({\partial\over\partial t } +{1\over t}D +
	\Delta^ {-1/2}F\Delta^{1/2}
\right)\Omega (t\vert x,x') = 0  \qquad ,
                                                               \eqno (11)
$$
$$
D=\sigma ^\mu \nabla _\mu
$$
 The initial and boundary conditions are to be determined from (3) and
(5)-(6).  But first let us discuss the general structure of this
construction.
As is generally known [1], the equation (8) determines the
geodetic interval defined as one half  the square of the length  of the
geodesic connecting the points $x$ and $x'$.  However in general case of
	topologically nontrivial manifolds there are more than one geodesic
	between points $x$ and $x'$, i.e. the equation (8) has more than one
	solution.
Therefore the quasi-classical ansatz, in general case,
	should have a form of a sum of analogous contributions from all
	geodesics. There is always one leading solution $\s (x,x')$
	which is determined by the shortest geodesic.
	It is marked out by the fact that it goes to zero
	$$ [\sigma ]\equiv\lim_{x\to x'}\s (x,x')=0
                                                            \eqno (12)
	$$
	when the points  $x$ and $x'$ approach each other.
	By the square brackets we denote here and below the coincidence
	limits of two-point quantities when the points
	 $x$ and $x'$ tend to each other along the shortest geodesic.
	 \par
	If one fixes a sufficiently small region including the points
	 $x$ and $x'$  (when they are close enough to each other)
	then only  this single solution is left.
	Therefore it is in some sense local and does not depend on the global
	structure of the manifold.
Multiple geodesics are closely
	associated with two reasons reflecting the essentially global
	(topological) aspects of the manifold. First, there could be
	manifolds with closed geodesics. In this case in addition to the
	shortest geodesic there are always geodesics that emanate from
	 point $x'$ , pass through the whole manifold one or several times and
	return to the point $x$.  Second, geodesics could be reflected from
	boundaries of the manifold one or more times.
	 \par
	In general case there could be infinitely large number of additional
	geodesics.
	They can be ordered according to the value of the geodetic interval.
	It is obvious that the more times the geodesic is reflected from
	boundaries or passes through the whole manifold the larger  the
	geodetic interval is. And according to (7) the value of the
        geodetic interval immediately determines the weight of the
	contribution of each geodesic in quasi-classical approximation.
	Let us mention that in heat kernel asymptotic expansion contribute
	only those geodesics for which the  geodetic interval could vanish.
	(Note that in the euclidean signature used in this paper all
	the geodetic intervals are non-negative according to definition.)
	There is only one such a geodesic among all additional ones, namely,
	the geodesic with one reflection from the boundary. It is evident
	that it has the minimal geodetic interval in comparison with all
	others. Moreover, if one fixes sufficiently narrow strip of the
	manifold near the boundary including points  $x$  and  $x'$
	then the geodesic with one reflection will be the only additional
	geodesic. In that sense the corresponding solution to the equation
	(8) is also local, i.e. it reflects the local properties of the
	boundary and does not depend on the global structure of the manifold.
	The corresponding geodetic interval      $\phi (x,x')$  in the
	coincidence limit $x=x'$ equals doubled square of the normal distance
	to the boundary and, therefore, vanishes on the boundary
	$$
	[\phi]=2 r^2 \qquad,\qquad [\phi]\bigg\vert_{\partial M}=0  \eqno (13)
	 $$
	 Therefore the corresponding solution contributes in the heat kernel
	asymptotic expansion but only on the boundary. All other solutions
	to the equation (8) are essentially global and corresponding geodetic
	intervals are strictly positive and do not vanish anywhere.
	So for the analysis of the heat kernel asymptotic expansion
	it is sufficient to restrict oneself to the local term and the term
	  with one reflection
	$$ U(t)=(4\p t)^{-t/2}\left(\exp\left(-{\s \over 2t}\right)\
	\Delta^{1/2}\Omega(t) +
	\exp\left(-{\phi\over 2t}\right)\Psi(t)\right)       \eqno (14)
	  $$
	 where $\Psi (t)$  is the corresponding transfer function. (Here it
	is convenient not to single out the pre-exponential factor, i.e the
	Van~Vleck-Morette determinant.)
	 The functions $\phi $ and  $\Psi $  satisfy  equations analogous to
	 (8)-(11)
        $$
        \phi ={1\over 2}\phi _\mu\phi^\mu=
        {1\over 2}\phi_{\mu^\prime}\phi^{\mu^\prime}     \qquad ,
                                                            \eqno (15)
        $$
        $$
        \phi_\mu=\nabla_\mu\phi\ \ \ ,\ \ \
        \phi_{\mu'}=\nabla_{\mu'}\phi
        $$
	$$
	L\Psi=0
	$$
	$$
L={\partial\over\partial t }
	+{1\over t}\left(\phi^\m\nabla_\m
	+{1\over 2}\left(\phi^{;\m}_\m-d\right)\right) - F
	                                                   \eqno (16)
	$$
	Since the second term in (14) does not contribute in the limit
	$t\to 0$ outside of the boundary and remembering that
	 $[\Delta ]=1$ we get from (3) the initial condition for
	$\Omega$.
$$
\left[\Omega (0\vert  x,x^\prime )\right]= 1 \qquad .
                                                                \eqno (17)
$$
The boundary conditions (4) take now the form:\hb
	 Dirichlet
	$$
	\left(\Delta^{1/2}\Omega + \Psi\right)\bigg\vert _{x \in \partial M}=0
								\eqno (18)
	$$
	 Neumann
	$$
	\left(n^\m\nabla_\m + S\right)
	\left(\Delta^{1/2}\Omega + \Psi\right)\bigg\vert _{x \in \partial M}
	- {1\over 2t}\s_{,n}
	\left(\Delta^{1/2}\Omega - \Psi\right)\bigg\vert _{x \in \partial M}=0
	                                                         \eqno (19)
	$$
	where $\s_{,n}=n^\m\nabla_\m\s$.
	\par
	For calculating the function $\Omega$ it is sufficient to take
 advantage of the standard Schwinger-De Witt expansion
 $$
 \Omega(t)=\sum\limits_{k=0}^\infty {(-t)^k\over k!} a_k
                                                        \eqno (20)
 $$
 where   $a_k (x,x')$  are the so called Hadamard - Minakshisundaram -
 De Witt - Seeley coefficients (HMDS) which are determined completely
 independently of the boundary conditions by recursion relations
  $$
 \left(1+{1\over k}D\right)a_k\ =\ \Delta^{-1/2}F\Delta ^{1/2}a_{k-1}
                                                                  \eqno (21)
 $$
The initial conditions for these recursion relations are obtained from
   (17) and have the form
 $$
 [a_0] = 1
                                                                 \eqno (22)
 $$
To solve these relations we have
 elaborated in the papers [4] a special technique. It allows to calculate
 an arbitrary HMDS-coefficient in a  sufficiently effective way.
 It is shown there that the formal solution of the recursion relations (21)
        $$
        a_k= \left(1+{1\over k}D\right)^{-1}
	\Delta^ {-1/2}F\Delta^{1/2}
	        \left(1+{1\over k-1}
        D\right)^{-1}
	\Delta^ {-1/2}F\Delta^{1/2}
		\cdots(1+D)^{-1}
	\Delta^ {-1/2}F\Delta^{1/2}                             \eqno(23)
	$$
takes the practical meaning in the form of covariant Taylor series
         $$
         a_k={\h P}\sum_{n\ge 0}{{(-1)^n}\over {n!}}
            \sigma^{\mu'_1}\cdots\sigma
            ^{\mu'_n}\left[\nabla_{(\mu_1}\cdots\nabla_{\mu_n)}a_k\right]
                                                                   \eqno(24)
         $$
	where $\h P(x,x')$ is the parallel displacement operator along the
 shortest geodesic from point $x'$  to  the point $x$. In [4] the
	 coefficients of that series are calculated and the results of
	calculations up to $a_4$   are listed.
	That is why we will concentrate our attention here on
	the calculation of the
	function $\Psi$.
	\par
	Let us choose for convenience of further calculations a special
	coordinate system in the neighbourhood of the boundary
	$x^\m=(r,\theta^i)$, where  $r$ is the length of the geodesic arc
       normal to the boundary in the point $\theta$, the equation of
	the boundary  having the form $r=0$, and
	 $\theta^i$ are the normal riemannian coordinates on the boundary.
	 \par
	We assume all quantities to be analytic in the
	coincidence limit on the boundary $\theta = \theta^\prime$.
	Hence we will expand all
	quantities in covariant Taylor series in the neighbourhood of the
	boundary.
	 \par
	The further strategy is rather simple. One should introduce a small
	parameter reflecting the fact that parameter $t$ is small and
	the points  $x$ and $x'$ are close to each other and to the boundary
	$$
	t^{1/2}\sim r\sim r'\sim (\theta -\theta^\prime) \sim 	\varepsilon
	$$
	and construct a corresponding perturbation theory in this parameter.
	We expand the transfer operator $L$ (16) in a formal series in
	the small parameter $\varepsilon$
	$$
	L\left(\eps^2 t\vert\eps r,\eps r',\eps (\theta -\theta^{\prime}),
	\theta^{\prime}\right)
	= {1\over
	\eps^2}L_{-2} + {1\over \eps}L_{-1} + L_0 + \cdots
	                                                      \eqno (25)
	$$
	and seek for a solution to this equation of the form
	$$
	\Psi\left(\eps^2
       t\vert\eps r,\eps r',\eps (\theta -\theta^{\prime}),
	\theta^{\prime}\right) =
       \sum_{n\ge 0}
\eps^n\Psi_n\left(t\vert r,r',(\theta-\theta^\prime),\theta^\prime\right)
                                                                \eqno (26)
	 $$
	From the transfer equation we get easy recursion differential
	relations
	$$
        \eqalignno{
	L_{-2}\Psi_0 &= 0                          & (27a) \cr
	L_{-2}\Psi_1 &= -L_{-1}\Psi_0              & (27b)   \cr
	L_{-2}\Psi_2 &= -L_{-1}\Psi_1 - L_0\Psi_0  & (27c)      \cr}
	$$
	etc..
	\par
	One shows easily that the coefficients $\Psi_n$ satisfy the
	scaling properties
	$$
       \Psi_n\left(\eps^2 t\vert\eps r,\eps r',\eps (\theta
	-\theta^{\prime}), \theta^{\prime}\right) = \eps^n
       \Psi_n\left(t\vert r, r',(\theta
	-\theta^{\prime}), \theta^{\prime}\right)
                                                	\eqno (28)
	$$
	Analogous equations take place also for the function $\Omega$.
	Whereas these relations are sufficient for the calculation of
	all $\Omega_n$ , for single-valued calculation of $\Psi_n$ it is
	necessary to use additionally the boundary conditions (18) (or (19)).
	\par
	The main difference between them is that
	$\Omega_n$ are analytic in all variables while $\Psi_n$ are analytic
   in $(\theta-\theta^\prime)$ but are complicated functions of the variables
	 $t$, $r$ and $r'$ (one can show that they are analytic in the
	variables  $R=(r+r')$ and $u=(r-r')/(r+r')$, so that the point
	$r=r'=0$  is  singular).
	 \par
	Making use of the scaling property (28) one can get
	from (26) an important representation for $\Psi(t)$
$$
       \Psi\left(t\vert r,r',(\theta -\theta^\prime),
       \theta^{\prime}\right) = \sum_{n\ge 0}
       t^{n/2}
     \Psi_n\left(1\bigg\vert {r\over \sqrt t}, {r'\over \sqrt
       t},{{(\theta-\theta^\prime)}\over \sqrt t}, \theta^\prime\right)
	                                                    \eqno (29)
	 $$
	  Using (11) and (14) we write down
	 for the trace of the heat kernel
	  $$
	 \eqalignno{
	 Tr\,[f\,U(t)] &= \int_{M} dx\,g^{1/2}tr[f\,U(t)]=  & (30)\cr
	          &= (4\p t)^{-d/2}tr\left\{\int_{M} dx\,g^{1/2}
         [f\,\Omega(t)]
	  +\int_{\partial M} d\theta\,\g^{1/2}\int^\infty_0 dr
          \exp\left({-r^2\over t}\right)
 {g^{1/2}\over \g^{1/2}}[f\,\Psi(t)] \right\} &
	\cr}
	  $$
     where $f(x)$ is an arbitrary smooth function on the manifold and
$\g = \det \g_{ij}, \g_{ij}  $ is the induced metric on the boundary.
	 One should mention that additional terms due to the presence of the
boundary contribute to the asymptotic expansion of the trace of the heat
kernel at  $t \to 0$ only when integrating over a sufficiently narrow
	 neighbourhood of the boundary. Therefore, the upper limit of the
	 integration over the distance to the boundary  $r$ in (30)
	 is taken to be $\infty $  (up to  exponentially small terms).
  \par
 At last, using (20) and (29) we obtain the asymptotic expansion of the
  trace of the heat kernel both in volume part and in surface one)
 $$
  \eqalignno{ TrfU(t)&=(4\p t)^{-d/2}tr\left\{\int_{M}dxg^{1/2}
\sum\limits_{k=0}^\infty{(-t)^k\over k!}[f\,a_k]
+\int_{\partial M}d\theta\g^{1/2}\sum\limits_{k=0}^\infty
	t^{{k+1}\over 2}c_{{k+1}\over 2}(f)\right\}
	                                                  & (31)\cr}
	$$
	 where
	 $$
          c_{{k+1}\over 2}(f) = \sum_{0\leq n \leq k}
\int^\infty_0 d\xi \exp\left(-{\xi^2}\right)\xi^n b_{k-n}(\xi)
\sum_{0\leq m \leq n}{g_m\over m!}{f^{(n-m)}\over (n-m)!}
							   \eqno (32a)
  $$
     $$
       b_k(\xi) = t^{-k/2}\left[\Psi_k\right]\bigg\vert_{r=\xi\sqrt t }
						       \eqno (32b)
$$
$$
f^{(k)} = n^{\m_1}\cdots n^{\m_k}\nabla_{\m_1}\cdots \nabla_{\m_k} f
\big\vert_{r=0}
$$
 $$
 g_m = \left({\partial \over \partial r}\right)^m\left({g^{1/2}\over
 \g^{1/2}} \right)\bigg\vert_{r=0}                         \eqno (33)
 $$
  Thus for the calculation of the boundary contributions in heat kernel
	asymptotic expansion one suffices to compute the coincidence limits
     $[\Psi_n]$,  put   $t=1$, integrate with the weight
	 $\exp(-r^2)r^m$ and  combine them with the quantities  $g_k$  (33)
	 calculated before.
  \bigskip
  \bigskip
  \bigskip
 \centerline{\bf 3.  Explicit expressions }
 \bigskip
Here we list the results of calculations of the lower order coefficients
of the heat kernel asymptotic expansion omitting most of the cumbersome
computations. The most complete list of volume contributions $a_k$ is
presented in [4]. The simplest first of them have the form
 $$
 \eqalignno{
 [a_0] &=1                 & \cr
 [a_1] &=Q-{{1}\over {6}} R                       & (34)\cr
 [a_2] &=\left(Q-{{1}\over {6}} R\right)^2 - {1\over
	3}\Square Q - {{1}\over {90}} R_{\mu\nu}R^{\mu\nu}
 + {{1}\over {90}}
	       R_{\mu\nu\alpha\beta}R^{\mu\nu\alpha\beta} + {{1}\over
 {15}}\Square R + {{1}\over {6}}\h R_{\mu\nu}\h R^{\mu\nu }   & \cr}
 $$
 where
  $\h  R_{\m\n} = [\nabla_\m,\nabla_n]$.
 \par
 Using the set of normal coordinates introduced above it is not difficult
to calculate the coefficients $g_k$  (33)
  $$
 \eqalignno{
 g_0 &= 1  & \cr
 g_1 &= -K & \cr
 g_2 &= -K_{ij}K^{ij} + K^2 - R^0_{nn}                    & (35) \cr}
 $$
  where
	$$
 \eqalignno{
 K &= \g^{ij} K_{ij}  & \cr
 K_{ij} &= -{1\over 2} {\partial \over \partial r} g_{ij}\bigg\vert_{r=0}& \cr
 \g_{ij} &= g_{ij}\bigg\vert_{r=0}                   & \cr
	R^0_{nn} &= n^\m n^\n R_{\m\n}\bigg\vert_{r=0}  & \cr}
	$$
    For simplicity we list below the solution only for Dirichlet boundary
conditions. From equation (27) and boundary condition (18) we have
	$$
	\Psi_0 = -1                      \eqno (36)
	$$
	The calculation of next orders is considerably more difficult
though it offers no particular problems. The result has the form
	$$
 \eqalignno{
	[\Psi_1] &=
	\sqrt t \left\{-{r^2\over t}
       h\left({r\over \sqrt t}\right) K	\right\} \cr
	[\Psi_2] &=
	t \Biggl\{ \left(Q - {1\over 6}\hat R\right) -
	{1\over 3}\left(1 + {r^2 \over t}\right) R^0_{nn}  \cr
	&+ f_1\left({r\over 	\sqrt t}\right) K^2
	+ f_2\left({r\over \sqrt t}\right)K_{ij}K^{ij} \Biggr\} & (37) \cr}
	$$
	where all tensor quantities are calculated on the boundary and
 $\hat R$ is the scalar curvature of the boundary
	$$
	\eqalignno{
          h(z) &= \int^\infty_0 dx \exp(-x^2 -2zx)= \exp(z^2) Erfc(z) &\cr
	 f_1(z) &= {1\over 6} + {z^2\over 6}\left( 2 + {1\over 2} z^2 -
	z(z^2 + 6) h(z)\right)                   & \cr
	f_2(z) &= -{1\over 6} +{1\over 12}z^2 \left( -4 + {1\over 2}z^2 -
	4z^3 h(z)\right)
                                                             & (38) \cr}
	$$
	Using these expressions and the integrals
	$$
	\int^\infty_0\exp(-\xi^2)\xi^n\,h(\xi)
	= {\Gamma\left({n+2\over 2}\right)\over 2(n+1)}
	$$
	we get finally several first boundary coefficients in asymptotic
expansion (31)
	$$
	\eqalignno{
        c_{1/2}(f) &= -{\sqrt \p	\over2}      & \cr
         c_1(f)    &= {1\over 3}K - {1\over 2}f^{(1)}& \cr
        c_{3/2}(f) &= {\sqrt \p\over 2}\Biggl\{
\left( -{1\over 6}\hat R - {1\over
		4}R^0_{nn} + {3\over 32} K^2 - {1\over 16} K_{ij}K^{ij} +
		Q\right) f                            & \cr
	           &+ {5\over 16}K\,f^{(1)} - {1\over 4}f^{(2)}\Biggr\}
                                     & (39)\cr}
	$$
	These results coincide with ones obtained by completely independent
methods in [6] that confirms that the approach developed in this paper
	is correct.
	\par
	Obtained results may be used when investigating Green function and
	the energy-momentum tensor of quantum fields near the boundary.
	\bigskip
       \bigskip
       \bigskip
	\centerline{\bf	3. Conclusion }
	\bigskip
	In this paper a new method for calculating the heat kernel asymptotic
	expansion (or one-loop effective action in any model of field theory)
	in the case of arbitrary background fields and manifolds of nontrivial
	topology (with boundary) is proposed. This method is a generalization
	of our previously elaborated  covariant technique for calculation of
	the effective action and is based on the summation in the
	quasi-classical  approximation the contributions of all geodesics
	connecting any two  points of the manifold (including the
	geodesics reflected from the boundary). It is established
	 that taking into
	account one additional geodesic with single reflection correctly
	reproduces all boundary contributions in the  heat kernel asymptotic
	expansion. Proposed approach allows not only to calculate
	the asymptotic expansion of the {\it trace} of the heat kernel but
	also to analyze the
	 {\it local} structure of the heat kernel near the
	boundary. Furhtermore we are going to calculate by means of it also
	the next terms of the asymptotic expansion (31) $\sim t^2$ and $\sim
 t^{5/2}$, i.e. the coefficients $c_2$ and $c_{5/2}$ (32).

\bigskip
\bigskip
\centerline{\bf References}
\bigskip
\item{1.} B. S. De Witt,
          Dynamical theory of groups and  fields
          (Gordon and Breach, New York,  1965)
\item{ 2.}N. D. Birrell and P. C. W. Davies,
          Quantum fields  in  curved space
          (Cambridge University Press, Cambridge, 1982)
\item{ 3.}A. O. Barvinsky and G. A.  Vilkovisky,
          Phys. Rep.   119 (1985) 1
\item{ 4.}I. G. Avramidi, Nucl. Phys.   B355 (1991) 712
\item{   }I. G. Avramidi, Phys. Lett.   B238 (1990) 92
\item{   }I. G. Avramidi, Phys. Lett.   B236 (1990) 443
\item{   }I. G. Avramidi, Teor. Mat. Fiz. 79 (1989) 219
\item{   }I. G. Avramidi, Yad. Fiz.       49 (1989) 1185
\item{ 5.}P. B.  Gilkey,
          Invariance theory,  the  heat  equation  and the
	  Atiyah-Singer index theorem
          (Publish or Perish, Wilmington, DE, 1984)
\item{ 6.}T. P.  Branson  and 	P.  B.  Gilkey,
          The  asymptotics  of  the laplacian on a manifold with  boundary,
          Odense  Universitet preprint No 17 (1989)
\item{ }  P. B. Gilkey, Comm. Part. Diff. Eq.         15 (1990) 245
\item{ 7.}H. P. McKean and I. M. Singer, J Diff. Geom. 1 (1967) 43
\item{ 8.}J. Melmed, J. Phys. A: Math. Gen.           21 (1989) L1131
\item{   }I. G. Moss, Class. Quant. Grav.              6 (1989) 759
\item{   }I. G. Moss and J. S. Dowker, Phys. Lett.  B229 (1989) 261
\item{   }J. S. Dowker and J. P. Schofield, J. Math. Phys. 31 (1990) 808
\item{ 9.}R. Balian and C. Bloch, Ann. Phys.           60 (1970) 401
\item{   }R. Balian and C. Bloch, Ann. Phys.           64 (1971) 271
\item{   }R. Balian and C. Bloch, Ann. Phys.           69 (1972) 76
\item{10.}G. Kennedy, J. Phys. A.: Math. Gen.          11 (1978) L173
\item{   }G. Kennedy, R. Critchley and  J. S. Dowker, Ann. Phys. 125 (1980) 346
\item{   }G. Kennedy, Ann. Phys.                               138 (1982) 353
\item{11.}R. T. Seeley, Am. Math. Soc. Prog. Pure Math.          10 (1967) 172
\item{   }M. Atiyah, R. Bott and V. K. Patodi, Invent. Math.    19 (1973) 279
\item{   }G. Cognola, L. Vanzo and S. Zerbini, Phys.  Lett.  B241  (1990) 381
\item{12} D. M. Mc Avity and H. Osborn, Class. Quant. Grav.      8 (1991) 603

\vfill
\eject
\end